\documentclass[aps,prb,amsmath,amssymb,amstext,citeautoscript,punctuation,nofootinbib,superscriptaddress,twocolumn]{revtex4-2}
\usepackage{amsthm,mathrsfs,amsfonts,dsfont,hyperref,epsfig,braket,bm,enumerate,color,graphicx,dcolumn,charter,gensymb,comment,physics,cleveref,mathtools,natbib}
\usepackage[normalem]{ulem}
\usepackage[dvipsnames]{xcolor}
\usepackage [english]{babel}
\newcommand{\ecs}{EuCd$_2$Sb$_2$}
\newcommand{\ezs}{EuZn$_2$Sb$_2$}
\begin{document}
\title{Magnetic structure of \ezs{} single-crystal thin-film} 
\author{Yu Wei Soh}
\affiliation{Quantum Innovation Centre (Q.InC), Agency for Science, Technology and Research (A*STAR), 2 Fusionopolis Way, Innovis \#08-03, Singapore 138634, Singapore}
\affiliation{Blackett Laboratory, Imperial College London, South Kensington, SW7 2AZ, London, United Kingdom}
\author{Hsiang Lee}
\affiliation{Department of Physics, Tokyo Institute of Technology, Tokyo 152-8551, Japan}
\author{Eugen Weschke}
\affiliation{Helmholtz-Zentrum Berlin für Materialien und Energie, Wilhelm-Conrad-Röntgen-Campus BESSY II, Albert-Einstein-Strasse 15, 12489 Berlin, Germany}
\author{Shinichi Nishihaya}
\affiliation{Department of Physics, Tokyo Institute of Technology, Tokyo 152-8551, Japan}
\author{Mikhael T. Sayat}
\affiliation{Quantum Innovation Centre (Q.InC), Agency for Science, Technology and Research (A*STAR), 2 Fusionopolis Way, Innovis \#08-03, Singapore 138634, Singapore}
\affiliation{Centre for Quantum Technologies, National University of Singapore, 3 Science Drive 2, Singapore 117543, Singapore}
\author{Masaki Uchida}
\affiliation{Department of Physics, Tokyo Institute of Technology, Tokyo 152-8551, Japan}
\affiliation{Toyota Physical and Chemical Research Institute (TPCRI), 480-1192 Nagakute, Japan}
\author{Jian-Rui Soh}
\affiliation{Quantum Innovation Centre (Q.InC), Agency for Science, Technology and Research (A*STAR), 2 Fusionopolis Way, Innovis \#08-03, Singapore 138634, Singapore}
\affiliation{Centre for Quantum Technologies, National University of Singapore, 3 Science Drive 2, Singapore 117543, Singapore}
\affiliation{Research School of Physics, Australian National University, Canberra, ACT, 0200, Australia}
\begin{abstract}
Magnetic topological materials are a class of compounds which can host massless electrons controlled by the magnetic order. One such compound is \ezs{}, which has recently garnered interest due to its strong interplay between the Eu magnetism and charge carriers. However the topology of the electronic band structure, which depends on the ground state magnetic configuration of the europium sublattice, has not been determined.  Based on our \textit{ab-initio} calculations, we find that an in-plane and out-of-plane \textit{A}-type antiferromagnetic (AFM) order generates a topological crystalline insulator and Dirac semimetal respectively, whereas a ferromagnetic (FM) order stabilizes a Weyl semimetal. Our resonant x-ray elastic scattering measurements of single-crystal thin film \ezs{} reveal both a sharp magnetic peak at $\textit{\textbf{Q}}$=$(0,0,\frac{1}{2})$ and broad $\textit{\textbf{Q}}$=$(0,0,1)$ below $T_{\mathrm{N}}=12.9$\,K, which is associated with an \textit{A}-type AFM and FM order, respectively. Our measurements indicate that the FM and AFM layers are spatially separated along the crystal $c$ axis, with the former limited to the top three atomic layers. We propose that \ezs{} behaves as a Weyl semimetal in the surface FM layers, and as a topological crystalline insulator in the lower AFM layers. 
\end{abstract}
\maketitle

\section{Introduction}

Magnetic topological materials is an emerging area of research, providing compelling analogies to particle physics models, along with technological applications~\cite{RevModPhys.90.015001}. A key feature is the ability to tune the electronic band topology via the magnetic order, which in turn can be modified by the coupling to an external magnetic field~\cite{wang_intrinsic_2023}. Materials which display a particularly strong coupling between magnetism and charge transport lend themselves well to exciting properties such as the control of massless Weyl or Dirac fermions, with potential applications in coherent spin transport~\cite{bernevig_progress_2022}. 

\begin{figure}[b!]
    \centering
\includegraphics[width=0.48\textwidth]{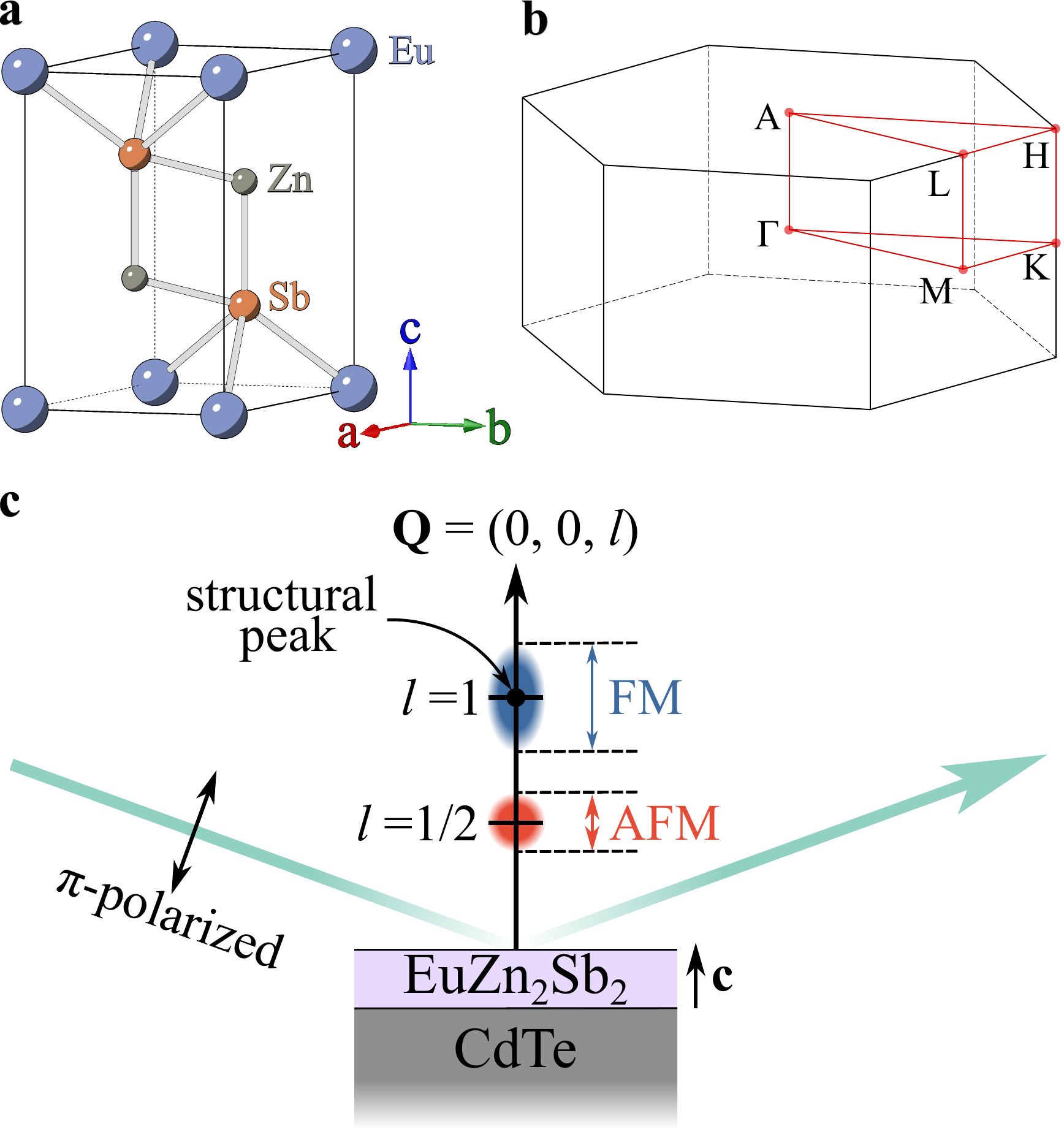}
    \caption{\textbf{a} Trigonal crystal structure of \ezs{}, showing $P\bar{3}m1$ space group. \textbf{b} The corresponding bulk Brillouin zone of \ezs{}. Red lines indicate the irreducible Brillouin zone with high symmetry points labelled. \textbf{c} Schematic diagram of REXS set-up, showing the incident and reflected x-ray (green arrow), and Bragg reflection peak of the structure (black), FM order (blue), and AFM order (red).  \label{fig:Crystal_Structure}}
\end{figure}

One such example is \ezs{} which belongs to a family of ternary Eu pnictides (Eu$X_2Y_2$, $X$=Cd, Zn, In; $Y$=Sb, As, P) and has attracted recent interest due to its strong interplay between the Eu magnetic order and electronic structure~\cite{Schellenberg2010,Schellenberg2011,HAMEED2025112519, riberolles_magnetic_2021, soh_understanding_2023, PhysRevB.104.174427, PhysRevB.108.054402}. This coupling arises because of the trigonal ($P\bar{3}m1$) crystal structure of \ezs{}, where the Zn-Sb charge transport layers are sandwiched between Eu magnetic layers~\cite{Schellenberg2010}, along the crystal $c$ axis [Fig.~\ref{fig:Crystal_Structure}\textbf{a}]. However, the ground state magnetic order and its influence on the topology of the electronic band structure has not yet been determined. 

\begin{figure*}[t!]
\centering
\includegraphics[width= 0.98\textwidth]{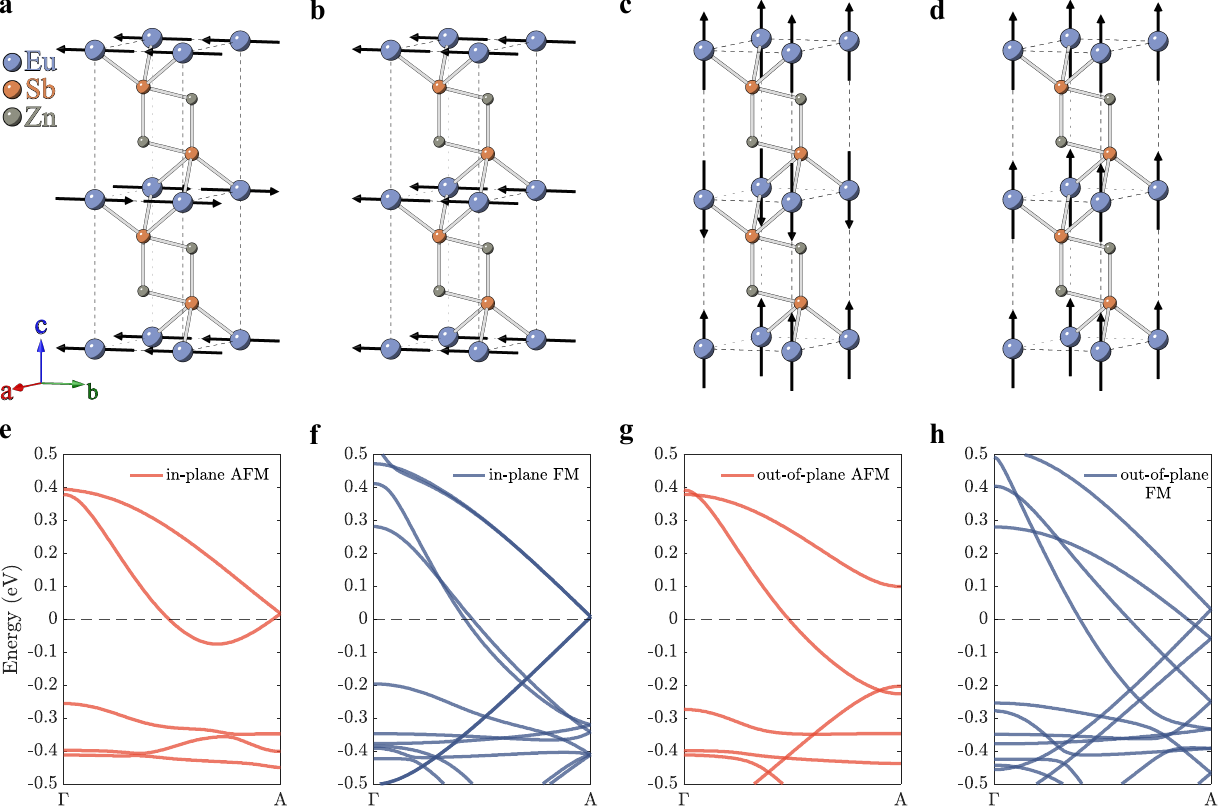}
\caption{ (\textbf{a}--\textbf{d}) Possible magnetic structures of \ezs{} with the Eu$^{2+}$ magnetic moments represented by black arrows and (\textbf{e}--\textbf{h}) corresponding electronic band structure along $\Gamma - \mathrm{A}$ high symmetry line, where the dashed lines at 0\,eV denote the Fermi energy ($E_{\mathrm{F}}$).
Magnetic structure and corresponding band structure of \textbf{a}, \textbf{e} in-plane AFM with doubly-degenerate bands and gapped Dirac; \textbf{b}, \textbf{f} in-plane FM with Weyl nodes; \textbf{c},\textbf{g} out-of-plane AFM with doubly-degenerate bands and Dirac point; \textbf{d}, \textbf{h} out-of-plane FM with Weyl nodes. ~\label{fig:DFT}}
\end{figure*}

Prior bulk magnetization measurements have reported that \ezs{} display an AFM order below the Néel temperature ($T_{\mathrm{N}}$) of around 13\,K~\cite{PhysRevB.73.014427,10.1063/1.3001608,10.1063/1.3681817,PhysRevB.109.125107}. However, despite obtaining similar magnetic susceptibility curves, Weber \textit{et al.}.~\cite{PhysRevB.73.014427} and May \textit{et al.}~\cite{10.1063/1.3681817} arrived at different conclusions on the directions of Eu magnetic moments associated with the AFM order, with the former~\cite{PhysRevB.73.014427} arguing that the moments lie within the $a$-$b$ plane [Fig.~\ref{fig:DFT}\textbf{a}], while the latter~\cite{10.1063/1.3681817} proposing that the Eu moments are along the crystal $c$ axis [Fig.~\ref{fig:DFT}\textbf{c}]. 

More generally, it is not yet clear how the magnetic configuration of the Eu$^{2+}$ sublattice influences the electronic band topology of \ezs{}. While various studies have reported the electronic band structure calculations of \ezs{}~\cite{10.1063/1.3001608, PhysRevB.111.205127, PhysRevB.110.045130}, a comprehensive study of all symmetry-allowed magnetic configurations has not been considered. For example, Zhang \textit{et al.}~\cite{10.1063/1.3001608} neglects the effect of both the magnetism of Eu ions and localization of $4f$ electrons which led to an incorrect prediction of a metallic band structure; Li \textit{et al.}~\cite{PhysRevB.111.205127} neglects the possibility of FM order; Sprague \textit{et al.}~\cite{PhysRevB.110.045130} did not consider in-plane FM order.

This work aims to address these two open questions surrounding \ezs{}:

Firstly, using density functional theory (DFT), we computed the various band structures assuming different symmetry-allowed Eu$^{2+}$ magnetic configurations, and study the implications on the electronic band topology. From our \textit{ab-initio}  calculations, we find that in-plane and out-of-plane AFM ordering leads to a gapped Dirac topological crystalline insulator (TCI) and Dirac semimetal (DSM) respectively, while both in-plane and out-of-plane FM ordering leads to a Weyl semimetal (WSM). Crucially, our calculations shows that the magnetic order of the Eu$^{2+}$ sublattice determines the topological class of the electronic bands of \ezs{}, since WSMs are protected by topology whereas the TCI and DSM are not.  

Secondly, using resonant x-ray elastic scattering (REXS), we determined the Eu magnetic order of single-crystal thin-film \ezs{}. Our REXS measurements reveal that thin film \ezs{} possesses an \textit{A}-type AFM ordering, where the europium moments are ferroically ordered within the $a$-$b$ plane but are antiferroically ordered along the crystal $c$ axis. Furthermore, we find evidence for FM ordering on the top three layers of the \ezs{} sample, which is likely formed due to surface oxidation. These findings point to a coexistence of AFM and FM orderings.

Furthermore, in terms of the electronic band topology, this spatially-separated coexistence of FM and AFM orders also suggests that the top few layers of \ezs{} harbour massless Weyl fermions, while the bottom layers contain topologically trivial charge carriers. Our study underscores the importance of ascertaining the magnetic order of magnetic topological materials.

\section{Methods}

Single crystal \ezs{} $(001)$ thin film was grown on CdTe $(111)$A substrates via the molecular beam epitaxy (MBE) method as outlined in Ref.~\cite{PhysRevB.111.L241106}. This MBE growth follows a similar method for EuCd$_2$$X_2$ ($X$=As, Sb) films~\cite{10.1063/1.5129467, PhysRevB.105.L201101, 10.1063/5.0183907}. First, before \ezs{} growth, the oxide layer on CdTe substrates was removed by etching CdTe $(111)$A substrates using 0.01\% Br$_2$- methanol, followed by annealing at $\sim$$550^{\circ}\mathrm{C}$ using Cd flux. Next, \ezs{} thin films were grown at a temperature of $\sim$$265^{\circ}\mathrm{C}$. The flux ratio of Eu, Zn, Sb used was $P_\mathrm{Eu}$: $P_\mathrm{Zn}$: $P_\mathrm{Sb}$ = 1: 7 to 15 : 2 to 3 respectively to grow films with a target thickness of $50$\,nm. The flux ratio was adjusted to tune hole densities from $10^{19}$ to $10^{20}$ cm$^{-3}$. Finally, the growth resulted in a \ezs{} thin film with the crystal $c$ axis oriented parallel to the CdTe substrate surface normal. 

To understand the effects of the europium magnetic order on the electronic band topology of \ezs{}, we performed \textit{ab-initio} band structure calculations using DFT, similar to the method used for \ecs{}~\cite{PhysRevB.98.064419}. The calculation was implemented using the Quantum Espresso suite~\cite{Giannozzi_2009}. For the self-consistent calculations, we used a $4 \times 4 \times 6$ Monkhorst-Pack grid for sampling~\cite{PhysRevB.13.5188}. The exchange correlation functional was calculated using the generalised gradient approximation (GGA). First, to account for the magnetism of the Eu$^{2+}$ ions, a spin polarized calculation was used. Next, to account for the relativistic effects of spin-orbit coupling owing to the large atomic mass of Sb, relativistic pseudopotentials were adopted. Third, due to Eu being a rare-earth ion, the $4f$ electrons are highly localized and self-interaction cannot be neglected~\cite{PhysRevB.111.205127}. To treat the interactions between electrons, we applied a Hubbard-\textit{U} correction to the GGA functional. In this work, we used \textit{U} = $2.5$\,eV, which is selected based on ARPES measurements on \ezs{}~\cite{PhysRevB.110.045130}. This places the $4f$ electrons around $1.8$\,eV below the Fermi level.

To probe the magnetic order of thin film \ezs{}, we conducted REXS measurements at the XUV diffractometer endstation of UE46-PGM-1 located at the BESSY II synchrotron facility~\cite{Weschke_Schierle_2018}. A schematic of the experimental set up is shown in Fig.~\ref{fig:Crystal_Structure}\textbf{c}. The sample was cooled to $3.8$\,K using a helium flow cryostat so that the measurements will be in the magnetically ordered phase below $T_\mathrm{N}$. To take advantage of the resonant enhancement of the x-ray scattering arising from the magnetic $4f$ states of Eu$^{2+}$~\cite{Hill:sp0084, Fink_2013, PAOLASINI2008550}, we tuned the incident soft photon energy to the Eu $M_5$ edge at $1129.8$\,eV. 

The \ezs{} crystal was mounted with the $c$ axis within the scattering plane [Fig.~\ref{fig:Crystal_Structure}\textbf{c}], so as to access the scattered x-ray intensity along the $\textit{\textbf{Q}}$=$(0,0,l)$ direction in reciprocal space. The scattered x-rays were measured using an AXUV100 avalanche photodetector without polarization analysis. To suppress the signal due to charge scattering of the $\textit{\textbf{Q}}$=$(0,0,1)$ peak, an incident photon polarization of $\pi$ was used, since the associated scattering angle ($2 \theta$)  is close to 90$^{\circ}$. As such, the $\pi'$-polarized x-rays arising from charge scattering is suppressed, resulting in the scattered x-ray signal arising mostly from the Eu$^{2+}$ magnetic order~\cite{PAOLASINI2008550, Hill:sp0084}.

\section{Results}

\subsection{DFT Calculations} \label{sec:DFT}
To understand how the magnetic order alters the electronic band topology, we consider various possible magnetic configurations, namely, the in-plane AFM, out-of-plane AFM, in-plane FM, and out-of-plane FM [Figs.~\ref{fig:DFT}\textbf{a}--\textbf{d}]. The corresponding calculated electronic band dispersion along the $\Gamma$--A high symmetry direction are shown in Figs.~\ref{fig:DFT}\textbf{e}--\textbf{h}. For both the in-plane and out-of-plane AFM order [Figs.~\ref{fig:DFT}\textbf{a}, \textbf{c}], non-symmorphic time reversal symmetry (TRS) is preserved, which results in doubly-degenerate bands~\cite{PhysRevB.98.064419}. The AFM order with the europium moments pointing along the $c$ axis [Fig.~\ref{fig:DFT}\textbf{c}] also possess the $\mathrm{C}_3$ rotational symmetry along $z$, which protects the Dirac point along the $\Gamma - \mathrm{A}$ line from gapping out. As such,  doubly-degenerate valence and conduction bands meet at a four-fold degenerate Dirac point ($\sim$0.2\,eV below $E_\mathrm{F}$) and can support massless quasiparticle excitations [Fig.~\ref{fig:DFT}\textbf{g}]. On the other hand, an in-plane AFM order [Fig~\ref{fig:DFT}\textbf{a}] breaks this $\mathrm{C}_3$ rotational symmetry, resulting in the formation of an avoided Dirac crossing and a TCI~\cite{honma_antiferromagnetic_2023}. This is a similar conclusion reached by another computational study~\cite{PhysRevB.111.205127}. Crucially, the charge carriers associated with AFM order \ezs{}, be it arising from the Dirac semimetallic or TCI phases, do not enjoy the protection afforded by topology.

Conversely, in the case of FM order, the non-symmorphic TRS is broken, lifting the double-degeneracy that is otherwise present in the case with AFM order [See Figs.~\ref{fig:DFT}\textbf{f}, \textbf{h}]. This FM order leads to the formation of Weyl nodes which are protected by topology.

 In summary, we found that the electronic band topology depends strongly on the magnetic order of the Eu ions. Therefore, it is imperative to work out if \ezs{} has an AFM or FM order.
 
\subsection{REXS Measurements}

\begin{figure}[b!]
    \centering
\includegraphics[width=0.48\textwidth]{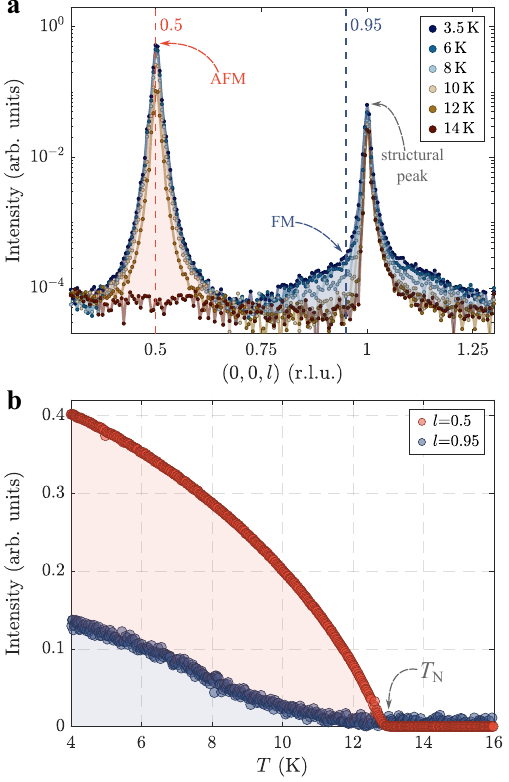}
    \caption{\textbf{a} REXS intensity in arbitrary units (arb. units) as a function of $(0, 0, l)$ at various temperatures, showing the emergence of peaks attributed to AFM order at $l = 0.5$ (red) and FM order at $l = 0.95$ (blue) from $12$\,K onwards. \textbf{b} REXS intensity as a function of temperature along the $l = 0.5$ and $l = 0.95$ line respectively, clearly showing the emergence of AFM order at $T_\mathrm{N} = 12.9$\,K and FM order at $T_\mathrm{C} \approx 13$\,K. \label{fig:REXS}}
\end{figure}

To ascertain whether \ezs{} possess an AFM or FM order, we turn to consider the REXS measurements of \ezs{} as shown in Fig.~\ref{fig:REXS}\textbf{a}. To probe the magnetic ordering along the crystal \textit{c} axis, we specifically consider the scattered x-ray intensity along $\textit{\textbf{Q}}$=$(0,0,l)$ direction obtained at various temperatures. The measurement at $T$=\,3.5\,K reveals a Bragg reflection peak at $\textit{\textbf{Q}}$=$(0,0,\frac{1}{2})$ which is otherwise structurally forbidden by the $P\bar{3}m1$ space group of \ezs{}. 

Figure~\ref{fig:REXS}\textbf{b} plots the temperature dependence of the $\textit{\textbf{Q}}$=$(0,0,\frac{1}{2})$ reflection, which shows the peak disappearing above $13$\,K, at the same temperature as the magnetic susceptibility peak seen in prior studies~\cite{PhysRevB.73.014427, 10.1063/1.3001608, PhysRevB.109.125107, 10.1063/1.3681817}. Given that the REXS was performed at the europium $M_5$ edge, the measured signal should be dominated by that arising from Eu$^{2+}$ ions. Moreover, the crystal structure of \ezs{} is not expected to change across $T_\mathrm{N}$ since Eu$^{2+}$ has no orbital angular momentum, leading to weak coupling of the magnetic order to the lattice. The reciprocal space location of the $\textit{\textbf{Q}}$=$(0,0,\frac{1}{2})$ peak suggests the formation of a magnetic order which doubles the unit cell in real space which is consistent with an \textit{A}-type AFM order, in accordance with that which has been reported in Refs.~\cite{PhysRevB.73.014427, 10.1063/1.3001608, PhysRevB.109.125107, 10.1063/1.3681817}. 

We now look for evidence for the FM correlations in \ezs{}, proposed by Sprague \textit{et al.}~\cite{PhysRevB.110.045130}. The REXS experimental signature of the FM order should arise at $\textit{\textbf{Q}}$=$(0,0,1)$ since the magnetic unit cell associated with a FM order has the same size as the structural unit cell of \ezs{}. However, the expected FM REXS signal might coincide with the structurally-allowed $\textit{\textbf{Q}}$=$(0,0,1)$ peak and hence be masked by the strong charge scattering. As described earlier in the methods section, we used $\pi$-polarized incident x-rays so that the charge scattering in the $\pi$$\rightarrow$$\pi'$ channel is suppressed while the magnetic scattering in the $\pi$$\rightarrow$$\pi'$ and $\pi$$\rightarrow$$\sigma'$ channels are not~\cite{PAOLASINI2008550, Hill:sp0084}.

As shown in Fig.~\ref{fig:REXS}\textbf{a}, the scattered x-ray intensity along $\textit{\textbf{Q}}$=$(0,0,l)$ measured at $T$=3.5K exhibits a narrow peak at $l = 1$ due to the leakage of the charge scattering and an additional broad signal centered at $\textit{\textbf{Q}}$=$(0,0,1)$, extending approximately from $l$$\sim$$0.70$ r.l.u. to $l$$\sim$$1.3$ r.l.u.. 

To determine the nature of this diffuse signal, we plot in Fig~\ref{fig:REXS}\textbf{b} the REXS signal at $l$=$0.95$ as a function of temperature. We chose $l$=$0.95$ to avoid the strong structural peak at $l$=$1$ but still to be sensitive to the diffuse magnetic signal. Interestingly, the diffuse FM signal also appears at $\sim$$13$\,K, concomitantly with the appearance of the AFM order at $T_\mathrm{N}$. The existence of both an AFM and FM peak below 13\,K indicate a surprising coexistence of magnetic orders. 

\begin{figure}[b!]
    \centering
\includegraphics[width=0.48\textwidth]{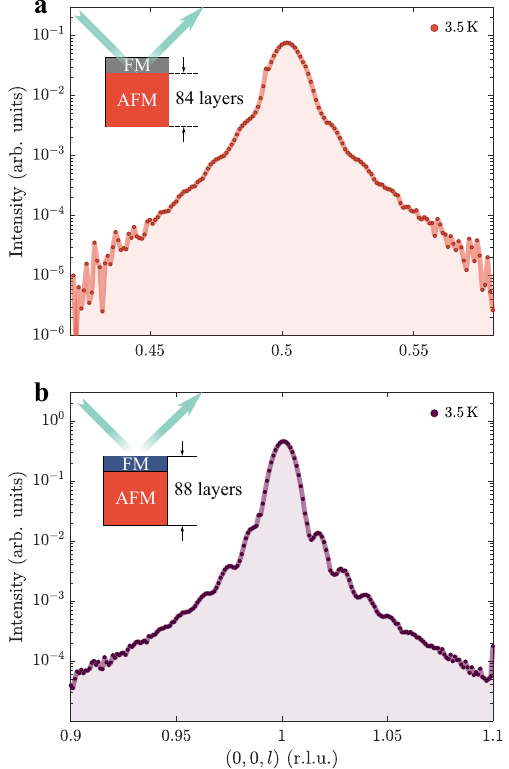}
    \caption{Laue oscillations of AFM peak at $(0, 0, \frac{1}{2})$ and structural peak at $(0, 0, 1)$, both measured at $3.5$\,K. Inset diagrams depict the thickness measured by the respective oscillations. \textbf{a} REXS intensity of $(0, 0, \frac{1}{2})$ AFM peak measured at the slightly off-resonance condition at $E_i = 1.128$\,keV, using $\pi$-polarized x-rays. This measures thickness of AFM layer. \textbf{b} Non-resonant x-ray scattering intensity of $(0, 0, 1)$ structural peak measured at $E_i = 1.121$\,keV, using $\sigma$-polarized x-rays. This measures total number of layers. \label{fig:Laue}}
\end{figure}

In order to understand how the AFM and FM order in \ezs{} can coexist with each other, we examine the associated thickness of the layers, to see if the two types of magnetic orders are spatially separated along the crystal $c$ axis. Firstly, we can estimate the number of FM layers from the low-temperature REXS measurements as shown in Fig~\ref{fig:REXS}\textbf{a}. Qualitatively, the diffuse FM scattering signal centered at $l$=$1$ is noticeably broader along $\textit{\textbf{Q}}$=$(0,0,l)$ as compared to the AFM signal at $l$=$\frac{1}{2}$, which suggests that in real space, there is a smaller number of layers with FM order as compared to the AFM order. A fit to the REXS intensity around $l$=$1$ shows that there are 3 layers associated with the FM order. 

Secondly, the thickness of the AFM layer can be estimated by measuring the Laue oscillations associated with the $\textit{\textbf{Q}}$=$(0, 0, \frac{1}{2})$ peak (See~\cite{Supplemental} for more details). To measure the thickness of the AFM order, the incident x-ray energy $E_\mathrm{i}$ was tuned to be slightly off-resonance from the europium $M_5$ edge to be sensitive to the magnetic order, but also reduce self absorption. Incident $\pi$-polarized x-rays were used to suppress charge scattering and enhance magnetic scattering as mentioned~\cite{Hill:sp0084}. Figure~\ref{fig:Laue}\textbf{a} plots the Laue oscillations at the $\textit{\textbf{Q}}$=$(0, 0, \frac{1}{2})$ peak, associated with the AFM order, at $3.5$\,K. Our fit to the measured signal indicates a thickness of 84 layers associated with AFM order. 

Finally, to measure the total thickness of the \ezs{} sample above the CdTe substrate, $E_\mathrm{i}$ was tuned to be off-resonance. Here, $\sigma$-polarized incident x-rays were used to mainly detect charge scattering arising from the structural peak~\cite{Hill:sp0084}. Figure~\ref{fig:Laue}\textbf{b} plots the Laue oscillations at the $\textit{\textbf{Q}}$=$(0, 0, 1)$ structural peak measured at $3.5$\,K. Our fit indicates approximately 88 layers corresponding to the total thickness of the \ezs{} sample. Interestingly, the total number of layers of \ezs{} is very close to the sum of AFM and FM layers, which suggest that the two different magnetic orders are spatially separated, along the crystal $c$ axis, which is also observed in EuCd$_2$Sb$_2$~\cite{PhysRevB.110.024405} a compound that is isostructural to \ezs{}. 

We now explain the mechanism that causes FM exchange coupling, and why this mechanism is only limited to a few layers of the \ezs{} crystal. The limited number of layers associated with FM order is possibly caused by surface oxidation, which limits the FM exchange coupling to the top few layers of the \ezs{} crystal. A similar mechanism has been reported for \ecs{}~\cite{PhysRevB.110.024405}. Firstly, surface oxidation creates a protective Eu$_2$O$_3$ layer~\cite{doi:10.1021/am5089007, doi:10.1021/acs.inorgchem.1c00708} that only affects the surface. Secondly, oxidation causes the conversion of Eu$^{2+}$ to Eu$^{3+}$. The mixed existence of Eu$^{2+}$ and Eu$^{3+}$ has also been reported for EuZnSb$_2$~\cite{PhysRevResearch.2.033462}. The conversion of Eu$^{2+}$ to Eu$^{3+}$ changes the sign of the exchange path between AFM ions along the $c$ axis, causing the ions to order ferromagnetically. Thus, surface oxidation explains the mechanism that only causes the top few layers to possess a FM ordering. 

To resolve the apparent disagreement of whether the AFM order should be in-plane~\cite{PhysRevB.73.014427} or out-of-plane~\cite{10.1063/1.3681817}, we refer to prior studies. Whilst our REXS measurements  do not resolve the direction of Eu$^{2+}$ magnetic moments for the AFM order, decisive experiments have been performed by Weber \textit{et al.}~\cite{PhysRevB.73.014427}, in which a spin-flop transition has been observed in the magnetization measurement with a field applied perpendicular to the crystal $c$ axis, at $B_{\mathrm{SF}} = 0.05$\,T. On the contrary, a spin-flop transition was not observed with the field applied along the crystal $c$ axis~\cite{PhysRevB.73.014427}, which suggest that the Eu$^{2+}$ moments associated with the AFM order lies in the $a$-$b$ plane. 

However, the out-of-plane orientation of magnetic moments associated with the AFM order proposed by May \textit{et al.} were based on magnetization measurements performed with field steps of $0.2$\,T along and perpendicular to the crystal $c$ axis~\cite{10.1063/1.3681817}. The step size of that in Ref.~\cite{10.1063/1.3681817} was too large to identify the spin-flop transition, which was observed by Weber \textit{et al.} to occur at very low fields of $B_\mathrm{SF}$=$0.05$\,T~\cite{PhysRevB.73.014427}. As such, we are more inclined to accept the conclusion by Ref.~\cite{PhysRevB.73.014427} that Eu$^{2+}$ ions have an in-plane AFM ordering. 

To decisively determine the orientation of the Eu$^{2+}$ magnetic moments associated with the AFM order, future REXS experiments should measure the dependence of the scattered intensity on (i) rotation of the sample about the scattering vector (azimuthal-angle dependence) and (ii) controlled variation of the incident linear polarization (full linear polarization analysis).

\section{Conclusion}
In this work, we successfully reconcile the existing debate on the magnetic ordering of \ezs{} and the effect on its electronic band topology. We first use DFT calculations to investigate how the magnetic ordering of \ezs{} determines the electronic band topology. The calculations show that an in-plane and out-of-plane AFM creates a gapped Dirac TCI and DSM respectively, while both in-plane and out-of-plane FM stabilizes a WSM. Next, the REXS measurements demonstrate the coexistence of an \textit{A}-type AFM and FM order. Notably, the FM order is confined to the top few layers, possibly due to surface oxidation. We also suggest a coupling of the AFM and FM orders, with the primary AFM order inducing the FM order. Finally, using the findings from DFT calculations and REXS, we predict that \ezs{} behaves as a WSM on the surface due to oxidation, and as a TCI on the lower unoxidised layers. Importantly, these belong to different topological classes.

\section*{Data Availability}
The data are not publicly available. The data are available from the authors upon reasonable request.

\begin{acknowledgments}
We acknowledge the BESSY II facility of the Helmholtz-Zentrum Berlin für Materialien und Energie for the provision of the beamtime under proposal No. 222-11565-ST. J.-R.S. and Y.W.S. acknowledges support from the Singapore National Science Scholarship, Agency for Science Technology and Research. This research is supported by A*STAR under Project No. C230917009, Q.InC Strategic Research and Translational Thrust; the MTC Young Investigator Research Grant (Award \# M24N8c0110); CQT++ Core Research Funding Grant (A*STAR).
\end{acknowledgments}
\bibliography{ref.bib}

@MISC{Supplemental,
   title        = {See {Supplemental Material
at [URL will be inserted by publisher] for details of the {Laue} oscillation model used to estimate the thickness of layers, which includes Refs.~\cite{Miller2022ExtractingIF}}}
}

@article{PhysRevB.73.014427,
  title = {Low-temperature properties and magnetic order of {$\mathrm{Eu}{\mathrm{Zn}}_{2}{\mathrm{Sb}}_{2}$}},
  author = {Weber, F. and Cosceev, A. and Drobnik, S. and Fai\ss{}t, A. and Grube, K. and Nateprov, A. and Pfleiderer, C. and Uhlarz, M. and L\"ohneysen, H. v.},
  journal = {Phys. Rev. B},
  volume = {73},
  issue = {1},
  pages = {014427},
  numpages = {7},
  year = {2006},
  month = {Jan},
  publisher = {American Physical Society},
  doi = {10.1103/PhysRevB.73.014427},
  url = {https://link.aps.org/doi/10.1103/PhysRevB.73.014427}
}

@article{10.1063/1.3001608,
    author = {Zhang, Hui and Zhao, Jing-Tai and Grin, Yu. and Wang, Xiao-Jun and Tang, Mei-Bo and Man, Zhen-Yong and Chen, Hao-Hong and Yang, Xin-Xin},
    title = {A new type of thermoelectric material, {EuZn$_2$Sb$_2$}},
    journal = {The Journal of Chemical Physics},
    volume = {129},
    number = {16},
    pages = {164713},
    year = {2008},
    month = {10},
    issn = {0021-9606},
    doi = {10.1063/1.3001608},
    url = {https://doi.org/10.1063/1.3001608}
}

@article{PhysRevB.109.125107,
  title = {Large unconventional anomalous Hall effect arising from spin chirality within domain walls of an antiferromagnet {${\mathrm{EuZn}}_{2}{\mathrm{Sb}}_{2}$}},
  author = {Singh, Karan and Pavlosiuk, Orest and Dan, Shovan and Kaczorowski, Dariusz and Wi\ifmmode \acute{s}\else \'{s}\fi{}niewski, Piotr},
  journal = {Phys. Rev. B},
  volume = {109},
  issue = {12},
  pages = {125107},
  numpages = {8},
  year = {2024},
  month = {Mar},
  publisher = {American Physical Society},
  doi = {10.1103/PhysRevB.109.125107},
  url = {https://link.aps.org/doi/10.1103/PhysRevB.109.125107}
}

@article{PhysRevB.110.045130,
  title = {Observation of paramagnetic spin-degeneracy lifting in {${\mathrm{EuZn}}_{2}{\mathrm{Sb}}_{2}$}},
  author = {Sprague, Milo X. and Regmi, Sabin and Ghosh, Barun and Sakhya, Anup Pradhan and Mondal, Mazharul Islam and Bin Elius, Iftakhar and Valadez, Nathan and Singh, Bahadur and Romanova, Tetiana and Kaczorowski, Dariusz and Bansil, Arun and Neupane, Madhab},
  journal = {Phys. Rev. B},
  volume = {110},
  issue = {4},
  pages = {045130},
  numpages = {6},
  year = {2024},
  month = {Jul},
  publisher = {American Physical Society},
  doi = {10.1103/PhysRevB.110.045130},
  url = {https://link.aps.org/doi/10.1103/PhysRevB.110.045130}
}

@article{doi:10.1021/am5089007,
author = {Averyanov, Dmitry V. and Sadofyev, Yuri G. and Tokmachev, Andrey M. and Primenko, Alexey E. and Likhachev, Igor A. and Storchak, Vyacheslav G.},
title = {Direct {Epitaxial Integration of the Ferromagnetic Semiconductor EuO with Silicon for Spintronic Applications}},
journal = {ACS Applied Materials \& Interfaces},
volume = {7},
number = {11},
pages = {6146-6152},
year = {2015},
doi = {10.1021/am5089007},
    note ={PMID: 25723051},
URL = {https://doi.org/10.1021/am5089007}
}

@article{doi:10.1021/acs.inorgchem.1c00708,
author = {Łażewski, Jan and Sternik, Małgorzata and Jochym, Paweł T. and Kalt, Jochen and Stankov, Svetoslav and Chumakov, Aleksandr I. and G{\"o}ttlicher, Jorg and R{\"u}ffer, Rudolf and Baumbach, Tilo and Piekarz, Przemysław},
title = {Lattice Dynamics and Structural Phase Transitions in {Eu$_2$O$_3$}},
journal = {Inorganic Chemistry},
volume = {60},
number = {13},
pages = {9571-9579},
year = {2021},
doi = {10.1021/acs.inorgchem.1c00708},
    note ={PMID: 34143607},
URL = {https://doi.org/10.1021/acs.inorgchem.1c00708}
}

@article{PhysRevResearch.2.033462,
  title = {Magnetic mixed valent semimetal {${\mathrm{EuZnSb}}_{2}$} with {Dirac} states in the band structure},
  author = {Wang, Aifeng and Baranets, Sviatoslav and Liu, Yu and Tong, Xiao and Stavitski, E. and Zhang, Jing and Chai, Yisheng and Yin, Wei-Guo and Bobev, Svilen and Petrovic, C.},
  journal = {Phys. Rev. Res.},
  volume = {2},
  issue = {3},
  pages = {033462},
  numpages = {8},
  year = {2020},
  month = {Sep},
  publisher = {American Physical Society},
  doi = {10.1103/PhysRevResearch.2.033462},
  url = {https://link.aps.org/doi/10.1103/PhysRevResearch.2.033462}
}

@article{PhysRevB.111.205127,
  title = {Manipulation of topological phase transitions and the mechanism of magnetic interactions in {Eu-based Zintl}-phase materials},
  author = {Li, Bo-Xuan and Song, Ziyin and Fang, Zhong and Wang, Zhijun and Weng, Hongming},
  journal = {Phys. Rev. B},
  volume = {111},
  issue = {20},
  pages = {205127},
  numpages = {12},
  year = {2025},
  month = {May},
  publisher = {American Physical Society},
  doi = {10.1103/PhysRevB.111.205127},
  url = {https://link.aps.org/doi/10.1103/PhysRevB.111.205127}
}

@article{PhysRevB.98.064419,
  title = {Magnetic and electronic structure of the layered rare-earth pnictide {${\mathrm{EuCd}}_{2}{\mathrm{Sb}}_{2}$}},
  author = {Soh, J.-R. and Donnerer, C. and Hughes, K. M. and Schierle, E. and Weschke, E. and Prabhakaran, D. and Boothroyd, A. T.},
  journal = {Phys. Rev. B},
  volume = {98},
  issue = {6},
  pages = {064419},
  numpages = {7},
  year = {2018},
  month = {Aug},
  publisher = {American Physical Society},
  doi = {10.1103/PhysRevB.98.064419},
  url = {https://link.aps.org/doi/10.1103/PhysRevB.98.064419}
}

@article{honma_antiferromagnetic_2023,
	title = {Antiferromagnetic topological insulator with selectively gapped {Dirac} cones},
	volume = {14},
	issn = {2041-1723},
	url = {https://doi.org/10.1038/s41467-023-42782-6},
	doi = {10.1038/s41467-023-42782-6},
	number = {1},
	journal = {Nature Communications},
	author = {Honma, A. and Takane, D. and Souma, S. and Yamauchi, K. and Wang, Y. and Nakayama, K. and Sugawara, K. and Kitamura, M. and Horiba, K. and Kumigashira, H. and Tanaka, K. and Kim, T. K. and Cacho, C. and Oguchi, T. and Takahashi, T. and Ando, Yoichi and Sato, T.},
	month = nov,
	year = {2023},
	pages = {7396},
}

@article{PhysRevB.111.L241106,
  title = {Difference in the in-plane anomalous Hall response in thin films of the Zintl compound {$\mathrm{Eu}{A}_{2}{\mathrm{Sb}}_{2}$} {($A=\mathrm{Zn},\mathrm{Cd}$)}},
  author = {Lee, Hsiang and Nishihaya, Shinichi and Kriener, Markus and Fujioka, Jun and Nakamura, Ayano and Watanabe, Yuto and Ishizuka, Hiroaki and Uchida, Masaki},
  journal = {Phys. Rev. B},
  volume = {111},
  issue = {24},
  pages = {L241106},
  numpages = {6},
  year = {2025},
  month = {Jun},
  publisher = {American Physical Society},
  doi = {10.1103/PhysRevB.111.L241106},
  url = {https://link.aps.org/doi/10.1103/PhysRevB.111.L241106}
}

@article{10.1063/1.5129467,
    author = {Su, Hao and Gong, Benchao and Shi, Wujun and Yang, Haifeng and Wang, Hongyuan and Xia, Wei and Yu, Zhenhai and Guo, Peng-Jie and Wang, Jinhua and Ding, Linchao and Xu, Liangcai and Li, Xiaokang and Wang, Xia and Zou, Zhiqiang and Yu, Na and Zhu, Zengwei and Chen, Yulin and Liu, Zhongkai and Liu, Kai and Li, Gang and Guo, Yanfeng},
    title = {Magnetic exchange induced {Weyl} state in a semimetal {EuCd$_2$Sb$_2$}},
    journal = {APL Materials},
    volume = {8},
    number = {1},
    pages = {011109},
    year = {2020},
    month = {01},
    issn = {2166-532X},
    doi = {10.1063/1.5129467},
    url = {https://doi.org/10.1063/1.5129467}
}

@article{PhysRevB.105.L201101,
  title = {Maximizing intrinsic anomalous Hall effect by controlling the {Fermi} level in simple {Weyl} semimetal films},
  author = {Ohno, Mizuki and Minami, Susumu and Nakazawa, Yusuke and Sato, Shin and Kriener, Markus and Arita, Ryotaro and Kawasaki, Masashi and Uchida, Masaki},
  journal = {Phys. Rev. B},
  volume = {105},
  issue = {20},
  pages = {L201101},
  numpages = {5},
  year = {2022},
  month = {May},
  publisher = {American Physical Society},
  doi = {10.1103/PhysRevB.105.L201101},
  url = {https://link.aps.org/doi/10.1103/PhysRevB.105.L201101}
}

@article{10.1063/5.0183907,
    author = {Nishihaya, Shinichi and Nakamura, Ayano and Ohno, Mizuki and Kriener, Markus and Watanabe, Yuto and Kawasaki, Masashi and Uchida, Masaki},
    title = {Intrinsic insulating transport characteristics in low-carrier density {EuCd$_2$As$_2$} films},
    journal = {Applied Physics Letters},
    volume = {124},
    number = {2},
    pages = {023103},
    year = {2024},
    month = {01},
    issn = {0003-6951},
    doi = {10.1063/5.0183907},
    url = {https://doi.org/10.1063/5.0183907}
}

@article{Giannozzi_2009,
doi = {10.1088/0953-8984/21/39/395502},
url = {https://dx.doi.org/10.1088/0953-8984/21/39/395502},
year = {2009},
month = {sep},
publisher = {},
volume = {21},
number = {39},
pages = {395502},
author = {Giannozzi, Paolo and Baroni, Stefano and Bonini, Nicola and Calandra, Matteo and Car, Roberto and Cavazzoni, Carlo and Ceresoli, Davide and Chiarotti, Guido L and Cococcioni, Matteo and Dabo, Ismaila and Dal Corso, Andrea and de Gironcoli, Stefano and Fabris, Stefano and Fratesi, Guido and Gebauer, Ralph and Gerstmann, Uwe and Gougoussis, Christos and Kokalj, Anton and Lazzeri, Michele and Martin-Samos, Layla and Marzari, Nicola and Mauri, Francesco and Mazzarello, Riccardo and Paolini, Stefano and Pasquarello, Alfredo and Paulatto, Lorenzo and Sbraccia, Carlo and Scandolo, Sandro and Sclauzero, Gabriele and Seitsonen, Ari P and Smogunov, Alexander and Umari, Paolo and Wentzcovitch, Renata M},
title = {QUANTUM ESPRESSO: a modular and open-source software project for quantum
simulations of materials},
journal = {Journal of Physics: Condensed Matter}
}

@article{PhysRevB.13.5188,
  title = {Special points for {Brillouin}-zone integrations},
  author = {Monkhorst, Hendrik J. and Pack, James D.},
  journal = {Phys. Rev. B},
  volume = {13},
  issue = {12},
  pages = {5188--5192},
  numpages = {0},
  year = {1976},
  month = {Jun},
  publisher = {American Physical Society},
  doi = {10.1103/PhysRevB.13.5188},
  url = {https://link.aps.org/doi/10.1103/PhysRevB.13.5188}
}

@article{PhysRevB.110.024405,
  title = {Magnetic structure of a single-crystal thin film of {${\mathrm{EuCd}}_{2}{\mathrm{Sb}}_{2}$}},
  author = {Heinrich, Eliot and Nakamura, Ayano and Nishihaya, Shinichi and Weschke, Eugen and R\o{}nnow, Henrik and Uchida, Masaki and Flebus, Benedetta and Soh, Jian-Rui},
  journal = {Phys. Rev. B},
  volume = {110},
  issue = {2},
  pages = {024405},
  numpages = {6},
  year = {2024},
  month = {Jul},
  publisher = {American Physical Society},
  doi = {10.1103/PhysRevB.110.024405},
  url = {https://link.aps.org/doi/10.1103/PhysRevB.110.024405}
}

@article{Weschke_Schierle_2018, 
title={The {UE46 PGM-1} beamline at {Bessy II}}, 
volume={4}, DOI={10.17815/jlsrf-4-77}, 
journal={Journal of large-scale research facilities JLSRF}, 
author={Weschke, Eugen and Schierle, Enrico}, 
year={2018}, 
month={Jan}}

@article{Hill:sp0084,
author = "Hill, J. P. and McMorrow, D. F.",
title = "{Resonant Exchange Scattering: Polarization Dependence and Correlation Function}",
journal = "Acta Crystallographica Section A",
year = "1996",
volume = "52",
number = "2",
pages = "236--244",
month = "Mar",
doi = {10.1107/S0108767395012670},
url = {https://doi.org/10.1107/S0108767395012670},
}

@article{bernevig_progress_2022,
	title = {Progress and prospects in magnetic topological materials},
	volume = {603},
	issn = {1476-4687},
	url = {https://doi.org/10.1038/s41586-021-04105-x},
	doi = {10.1038/s41586-021-04105-x},
	number = {7899},
	journal = {Nature},
	author = {Bernevig, B. Andrei and Felser, Claudia and Beidenkopf, Haim},
	month = mar,
	year = {2022},
	pages = {41--51},
}

@article{RevModPhys.90.015001,
  title = {Weyl and Dirac semimetals in three-dimensional solids},
  author = {Armitage, N. P. and Mele, E. J. and Vishwanath, Ashvin},
  journal = {Rev. Mod. Phys.},
  volume = {90},
  issue = {1},
  pages = {015001},
  numpages = {57},
  year = {2018},
  month = {Jan},
  publisher = {American Physical Society},
  doi = {10.1103/RevModPhys.90.015001},
  url = {https://link.aps.org/doi/10.1103/RevModPhys.90.015001}
}

@article{10.1063/1.3681817,
    author = {May, Andrew F. and McGuire, Michael A. and Ma, Jie and Delaire, Olivier and Huq, Ashfia and Custelcean, Radu},
    title = {Properties of single crystalline {AZn$_2$Sb$_2$ (A=Ca,Eu,Yb)}},
    journal = {Journal of Applied Physics},
    volume = {111},
    number = {3},
    pages = {033708},
    year = {2012},
    month = {02},
    issn = {0021-8979},
    doi = {10.1063/1.3681817},
    url = {https://doi.org/10.1063/1.3681817}
}

@article{wang_intrinsic_2023,
	title = {Intrinsic magnetic topological materials},
	volume = {18},
	issn = {2095-0470},
	url = {https://doi.org/10.1007/s11467-022-1250-6},
	doi = {10.1007/s11467-022-1250-6},
	number = {2},
	journal = {Frontiers of Physics},
	author = {Wang, Yuan and Zhang, Fayuan and Zeng, Meng and Sun, Hongyi and Hao, Zhanyang and Cai, Yongqing and Rong, Hongtao and Zhang, Chengcheng and Liu, Cai and Ma, Xiaoming and Wang, Le and Guo, Shu and Lin, Junhao and Liu, Qihang and Liu, Chang and Chen, Chaoyu},
	month = feb,
	year = {2023},
	pages = {21304},
}

@article{HAMEED2025112519,
title = {First-principles calculations to investigate magnetic, electronic, and thermoelectric response of europium-based half metallic ternary Zintl compounds {EuMg$_2$X$_2$} {(X=Sb and Bi)}},
journal = {Journal of Physics and Chemistry of Solids},
volume = {199},
pages = {112519},
year = {2025},
issn = {0022-3697},
doi = {https://doi.org/10.1016/j.jpcs.2024.112519},
url = {https://www.sciencedirect.com/science/article/pii/S0022369724006541},
author = {Uzma Hameed and Hayat Ullah and Syed Zeshan Abbas and Kashif Safeen and Khalid M. Alotaibi and Akif Safeen and Sadia Yasin and G. Murtaza and Fatima Khalil and Sajad Ali and Ghulam Asghar and Rajwali Khan},
}

@article{riberolles_magnetic_2021,
	title = {Magnetic crystalline-symmetry-protected axion electrodynamics and field-tunable unpinned {Dirac} cones in {EuIn$_2$As$_2$}},
	volume = {12},
	issn = {2041-1723},
	url = {https://doi.org/10.1038/s41467-021-21154-y},
	doi = {10.1038/s41467-021-21154-y},
	number = {1},
	journal = {Nature Communications},
	author = {Riberolles, S. X. M. and Trevisan, T. V. and Kuthanazhi, B. and Heitmann, T. W. and Ye, F. and Johnston, D. C. and Bud’ko, S. L. and Ryan, D. H. and Canfield, P. C. and Kreyssig, A. and Vishwanath, A. and McQueeney, R. J. and Wang, L. -L. and Orth, P. P. and Ueland, B. G.},
	month = feb,
	year = {2021},
	pages = {999},
}

@article{soh_understanding_2023,
	title = {Understanding unconventional magnetic order in a candidate axion insulator by resonant elastic x-ray scattering},
	volume = {14},
	issn = {2041-1723},
	url = {https://doi.org/10.1038/s41467-023-39138-5},
	doi = {10.1038/s41467-023-39138-5},
	number = {1},
	journal = {Nature Communications},
	author = {Soh, Jian-Rui and Bombardi, Alessandro and Mila, Frédéric and Rahn, Marein C. and Prabhakaran, Dharmalingam and Francoual, Sonia and Rønnow, Henrik M. and Boothroyd, Andrew T.},
	month = jun,
	year = {2023},
	pages = {3387},
}

@article{PhysRevB.104.174427,
  title = {{$A$}-type antiferromagnetic order and magnetic phase diagram of the trigonal {Eu spin-$\frac{7}{2}$} triangular-lattice compound {${\mathrm{EuSn}}_{2}{\mathrm{As}}_{2}$}},
  author = {Pakhira, Santanu and Tanatar, M. A. and Heitmann, Thomas and Vaknin, David and Johnston, D. C.},
  journal = {Phys. Rev. B},
  volume = {104},
  issue = {17},
  pages = {174427},
  numpages = {15},
  year = {2021},
  month = {Nov},
  publisher = {American Physical Society},
  doi = {10.1103/PhysRevB.104.174427},
  url = {https://link.aps.org/doi/10.1103/PhysRevB.104.174427}
}

@article{PhysRevB.108.054402,
  title = {Superexchange interaction in insulating {EuZn}$_2${P}$_2$},
  author = {Singh, Karan and Dan, Shovan and Ptok, A. and Zaleski, T. A. and Pavlosiuk, O. and Wi\ifmmode \acute{s}\else \'{s}\fi{}niewski, P. and Kaczorowski, D.},
  journal = {Phys. Rev. B},
  volume = {108},
  issue = {5},
  pages = {054402},
  numpages = {8},
  year = {2023},
  month = {Aug},
  publisher = {American Physical Society},
  doi = {10.1103/PhysRevB.108.054402},
  url = {https://link.aps.org/doi/10.1103/PhysRevB.108.054402}
}

@article{Fink_2013,
doi = {10.1088/0034-4885/76/5/056502},
url = {https://dx.doi.org/10.1088/0034-4885/76/5/056502},
year = {2013},
month = {apr},
publisher = {IOP Publishing},
volume = {76},
number = {5},
pages = {056502},
author = {Fink, J and Schierle, E and Weschke, E and Geck, J},
title = {Resonant elastic soft x-ray scattering},
journal = {Reports on Progress in Physics}
}

@article{PAOLASINI2008550,
title = {Magnetic and resonant X-ray scattering investigations of strongly correlated electron systems},
journal = {Comptes Rendus Physique},
volume = {9},
number = {5},
pages = {550-569},
year = {2008},
note = {Synchrotron x-rays and condensed matter},
issn = {1631-0705},
doi = {https://doi.org/10.1016/j.crhy.2007.06.005},
url = {https://www.sciencedirect.com/science/article/pii/S1631070507001569},
author = {Luigi Paolasini and François {de Bergevin}}
}

@article{Schellenberg2010,
author = {Schellenberg, Inga and Eul, Matthias and Hermes, Wilfried and Pöttgen, Rainer},
title = {A $^{121}${Sb} and $^{151}${Eu} {Mössbauer} {Spectroscopic} {Investigation} of {EuMn}$_2${Sb}$_2$, {EuZn}$_2${Sb}$_2$, {YbMn}$_2${Sb}$_2$, and {YbZn}$_2${Sb}$_2$},
journal = {Zeitschrift für anorganische und allgemeine Chemie},
volume = {636},
number = {1},
pages = {85-93},
doi = {https://doi.org/10.1002/zaac.200900413},
url = {https://onlinelibrary.wiley.com/doi/abs/10.1002/zaac.200900413},
year = {2010}
}

@article{Schellenberg2011,
author = {Schellenberg, Inga and Pfannenschmidt, Ulrike and Eul, Matthias and Schwickert, Christian and Pöttgen, Rainer},
title = {A $^{121}${Sb} and $^{151}${Eu} {Mössbauer} {Spectroscopic} {Investigation} of {EuCd}$_2${X}$_2$ ({X} = {P}, {As}, {Sb}) and {YbCd}$_2${Sb}$_2$},
journal = {Zeitschrift für anorganische und allgemeine Chemie},
volume = {637},
number = {12},
pages = {1863-1870},
doi = {https://doi.org/10.1002/zaac.201100179},
url = {https://onlinelibrary.wiley.com/doi/abs/10.1002/zaac.201100179},
year = {2011}
}
\end{document}